\documentclass[12pt]{iopart}
\usepackage{graphicx}
\usepackage{iopams}
\usepackage{wasysym}
\usepackage{color}
\usepackage[colorlinks=true, urlcolor=blue,citecolor=blue,linkcolor=blue]{hyperref}

\usepackage{amstext}
\usepackage{array}

\begin{document}

\title{Fidelity deviation in quantum teleportation}

\author{Jeongho Bang$^1$, Junghee Ryu$^2$, and Dagomir Kaszlikowski$^{2,3}$}
\address{$^1$ School of Computational Sciences, Korea Institute for Advanced Study, Seoul 02455, Korea}

\address{$^2$Centre for Quantum Technologies, National University of Singapore, 3 Science Drive 2, 117543 Singapore, Singapore}

\address{$^3$Department of Physics, National University of Singapore, 3 Science Drive 2, 117543 Singapore, Singapore}
\eads{\mailto{rjhui82@gmail.com}, \mailto{phykd@nus.edu.sg}}


\newcommand{\bra}[1]{\left<#1\right|}
\newcommand{\ket}[1]{\left|#1\right>}
\newcommand{\abs}[1]{\left|#1\right|}
\newcommand{\expt}[1]{\left<#1\right>}
\newcommand{\braket}[2]{\left<{#1}|{#2}\right>}
\newcommand{\ketbra}[2]{\left|{#1}\left>\right<{#2}\right|}
\newcommand{\commt}[2]{\left[{#1},{#2}\right]}
\newcommand{\openone}{\leavevmode\hbox{\small1\normalsize\kern-.33em1}}

\newcommand{\new}[1]{\textcolor{red}{#1}}

\begin{abstract}
We analyze the performance of quantum teleportation in terms of average fidelity and fidelity deviation. The average fidelity is defined as the average value of the fidelities over all possible input states and the fidelity deviation is their standard deviation, which is referred to as a concept of fluctuation or universality. In the analysis, we find the condition to optimize both measures under a noisy quantum channel---we here consider the so-called Werner channel. To characterize our results, we introduce a two-dimensional space defined by the aforementioned measures, in which the performance of the teleportation is represented as a point with the channel noise parameter. Through further analysis, we specify some regions drawn for different channel conditions, establishing the connection to the dissimilar contributions of the entanglement to the teleportation and the Bell inequality violation.
\end{abstract}
\maketitle

\section{Introduction}

Quantum teleportation, proposed by Bennett {\it et al}.~\cite{Bennett93}, is one of the fundamental quantum information protocols, which enables the remote transmission of an unknown quantum state via shared entanglement and classical communication. Many quantum teleportation protocols have been provided useful primitives for practical quantum technologies, e.g., rapid and secret quantum communications~\cite{Zukowski00,Barrett01,Cavalcanti13}. Furthermore, it offers a framework to study the interrelated principles of the entanglement, particularly establishing a link to those in Bell inequality violation \cite{Werner89,Popescu94}.

Fidelity $f$ (a closeness between initial and final states \cite{Jozsa94}) has widely been used to characterize the performance of various quantum information tasks. Note that $f$ is bounded by $0 \leq f \leq 1$, where the unit fidelity ($f=1$) implies that the initial and final states are equivalent. Quantum teleportation is designed to transmit all possible (unknown) input states, thus a measure of averaging over all inputs is used. That is the average fidelity $F$. It quantifies {\em how well} the unknown input states can be transmitted to another location. Thus, $F$ shows the optimality of the quantum teleportation. Many studies addressed the relation between the maximum achievable average fidelity and the amount of entanglement of the shared states~\cite{Horodecki99}. There it is shown that the unit average fidelity ($F=1$) can be obtained when the two remote parties share the maximally entangled states, whereas $F=2/3$ is the maximally attainable one in any classical schemes which cannot use the entanglement~\cite{Bennett93,Massar95,Gisin96}. 

However, the average fidelity tells nothing about {\em how equally} all possible input states are transmitted (unless $F$ is one or zero, we shall discuss later). Indeed, one cannot guarantee solely based on $F$ whether the given teleportation protocol performs equally to all input states. What we need is a measure of how far each value is spread apart. This is what the deviation can provide. In simple words, for the given teleportation protocol showing a certain average fidelity $F$, there might exists a specific input state (or a set of the states) that shows a very lower fidelity $f$ than the mean value $F$; namely, $f$ fluctuates on each input state. The investigation of such a fluctuation property will be quite critical, particularly when we use the teleportation scheme to implement element gates for universal computation \cite{Jeong02,Ralph03} or when unexpected noises and imperfect controls impinge on the protocols. It is thus desired to consider another measure to quantify the aforementioned in the teleportation.

The fidelity deviation $D$, introduced in \cite{Bang12}, can resolve the problem in question. It is defined by the standard deviation of the fidelities $f$ and is often referred to as the concept of fluctuation \cite{Pedersen08,Magesan11} or universality \cite{Bang12}. In what follows, we thus analyze the average fidelity and the fidelity deviation in the teleportation with a noisy channel. A family of two-qubit Werner state, which is a mixture of the maximally entangled state and white noise, is our example of the noisy channel. In the analysis, we present the condition to maximize the average fidelity $F$ and to minimize the fidelity deviation $D$. We then represent the performance of the teleportation as a point in the two-dimensional space defined by the measures $F$ and $D$. Through further analysis, we argue the dissimilar aspects of the entanglement contributions to the teleportation and the violation of Bell inequality \cite{Werner89, Popescu94}, specifying some point regions drawn for different channel conditions. In a recent paper~\cite{Cavalcanti17}, the quantum teleportation has been studied with full data available in an experiment, in which more information than the average fidelity is taken into account in analysis of teleportation scheme. This leads one to show that every entangled states can be used as a quantum channel in teleportation.



\section{Average fidelity and fidelity deviation}

Fidelity quantifies a transformation performance between an input state and its target state, which reads~\cite{Jozsa94}
\begin{eqnarray}
f = \Tr{\left[ \left(\sqrt{\hat{\tau}}\hat{\varrho}_\phi\sqrt{\hat{\tau}}\right)^{1/2}\right]},
\end{eqnarray}
where $\hat{\tau}$ is a density operator of the target state and $\hat{\varrho}_\phi$ is of the transformed state of the input state $\ket{\phi}$. The quantum teleportation applies to unknown input states so that the average fidelity---an average of the fidelities $f$ over all possible input states---is used:
\begin{eqnarray}
F=\int{d\phi}\,f,
\label{eq:avgf}
\end{eqnarray}
where $d\phi$ is Haar measure with $\int d\phi=1$. Here, $F=1$ implies that the task is perfectly performed for all possible inputs, while $F=1/2$ does at random.




Next, let us consider a situation that the fidelity $f$ fluctuates on the input states. To quantify such a fluctuation, we employ the fidelity deviation $D$, defined as the  standard deviation of $f$: 
\begin{eqnarray}
D = \sqrt{\int{d\phi}\, f^2 - F^2}.
\label{eq:stdd} 
\end{eqnarray}
Here, $D=0$ means no fluctuation of $f$, and this holds when $f=F$ for all input states. In such case, the task is said to be universal~\cite{Bang12}. Note that this fidelity deviation is bounded by $0 \le D \le 1/2$, which is obtained by
\begin{eqnarray}
D^2 \leq\int{d\phi}\,{f} - F^2=F(1-F)\leq\frac{1}{4},
\label{eq:limit_ug}
\end{eqnarray}
where the last inequality is saturated when $F=1/2$.

From the two formulas (\ref{eq:avgf}) and (\ref{eq:stdd}), one can see that the unit average fidelity $F=1$ holds if and only if $f=1$ for all possible inputs (or equivalently, $D=0$). However, when the attainable $F$ is limited to less than $1$, then the zero-fluctuation may not happen because the fidelity deviation $D$ can be in range from $0$ to $\sqrt{F(1-F)}$. For some probabilistic tasks, for example, universal-NOT \cite{Buzek99,Buzek00-1} and quantum cloning \cite{Buzek96}, this is a problem. Noting that in a realistic circumstance, it is difficult or even impractical to achieve $F=1$. Thus minimizing $D$ and maximizing $F$ are both demanded.

\section{Analysis of the two measures}\label{SEC:3}

Let us, briefly, recall the quantum teleportation protocol for a qubit case. A sender (say, Alice) and a receiver (say, Bob) share the maximally entangled state $\ket{\Psi_0} = \left(\ket{00} + \ket{11}\right)/\sqrt{2}$ as a quantum channel. Alice then performs joint (Bell) measurements on the states: one is to be teleported and the other is from the entangled pair. The measurement bases read
\begin{eqnarray}
\ket{\Psi_\alpha} = \hat{U}_\alpha \otimes \hat{\openone}_2 \ket{\Psi_0}, 
\end{eqnarray}
where $\alpha \in \{0,1,2,3\}$, $\hat{U}_\alpha$ is a single-qubit unitary, and $\hat{\openone}_2$ is identity operator in $2$-dimensional Hilbert space. After the measurement, Alice sends the outcome $m_\alpha$ to Bob through a classical channel, and Bob performs the corresponding unitary operation $\hat{V}_\alpha$ to his state (another one of the entangled pair). Then, the transformed state is to be the state that Alice wanted to send to Bob.


In a general teleportation scenario, one can consider a noisy quantum channel, that is, non-maximally entangled state is shared to the two parties. A family of Werner state is our working example of the channel, which is defined by a statistical mixture of the maximally entangled state $\ket{\Psi_0}$ with the white noise \cite{Werner89}:
\begin{eqnarray}
\hat{\varrho}_W = p \ket{\Psi_0}\bra{\Psi_0} + \frac{1}{4}\left(1-p\right)\hat{\openone}_4,
\label{eq:ws}
\end{eqnarray}
where $p$ is the noise parameter, bounded by $0 \le p \le 1$. It is known that the amount of the entanglement in the Werner state can be quantified by the noise parameter $p$; for example, the violation of Clauser-Horne-Shimony-Holt (CHSH) inequality \cite{Clauser69} is allowed when $p>1/\sqrt{2}$, and the inseparability condition of the Werner state is $p>1/3$ \cite{Werner89}. Also, the existence of local hidden variables (LHV) model for the Werner state is connected to the Grothendieck's constant, i.e., the critical value of allowing the LHV model is given by $1/K_{G}(3)$~\cite{Acin06}. Recent progress has been made in such values; $0.6829 \leq p \leq 0.6964$~\cite{Hirsch17, Divianszky17}. Later, we consider the critical values by the violation of CHSH inequality and inseparability condition in the analysis of the fidelity deviation.


Now, we shall analyze the average fidelity $F$ and the fidelity deviation $D$ in the general teleportation scenario. The finally teleported state $\hat{\varrho}_\phi$ reads
\begin{eqnarray}
\hat{\varrho}_\phi = \sum_{\alpha=0}^{3} \left(\hat{\openone}_{4} \otimes \hat{V}_\alpha\right) \bra{\Psi_\alpha} \hat{\varrho} \ket{\Psi_\alpha}  \left(\hat{\openone}_{4} \otimes \hat{V}_\alpha^\dagger\right),
\label{eq:out_st}
\end{eqnarray}
where $\hat{\varrho}=\ket{\phi}\bra{\phi} \otimes \hat{\varrho}_W$ is the total initial state with the quantum channel $\hat{\varrho}_W$ of (\ref{eq:ws}). The state (\ref{eq:out_st}) can be rewritten as
\begin{eqnarray}
\hat{\varrho}_\phi = \frac{p}{4}\sum_{\alpha=0}^{3}\hat{X}_\alpha \ket{\phi}\bra{\phi} \hat{X}_\alpha^\dagger + \frac{1}{2}\left(1-p\right)\hat{\openone}_2,
\label{eq:out_str}
\end{eqnarray}
where $\hat{X}_\alpha = \hat{V}_\alpha \hat{U}_\alpha^\dagger$. With the formula (\ref{eq:out_str}), the fidelity $f$ between the initial and final states reads
\begin{eqnarray}
f = \frac{p}{4}\sum_{\alpha=0}^{3}\xi_\alpha + \frac{1}{2}\left(1-p\right),
\label{eq:f_d}
\end{eqnarray}
where $\xi_\alpha$ is given by
\begin{eqnarray}
\xi_\alpha = \Tr{\left(\hat{X}_\alpha \ket{\phi}\bra{\phi} \hat{X}_\alpha^\dagger \ket{\phi}\bra{\phi}\right)} = \abs{\bra{\phi}\hat{X}_\alpha\ket{\phi}}^2.
\label{eq:df_xi}
\end{eqnarray}

\subsection{Average fidelity $F$}
In the Bloch representation, the initial state can be written by
\begin{eqnarray}
\ket{\phi}\bra{\phi} = \frac{1}{2}\left(\hat{\openone}_2 + \boldsymbol{\phi}^T \boldsymbol{\sigma}\right),
\label{eq:Bphi_st}
\end{eqnarray}
where $\boldsymbol{\phi} = (\phi_x, \phi_y, \phi_z)^T$ is called the Bloch vector and $\boldsymbol{\sigma} = (\hat{\sigma}_x, \hat{\sigma}_y, \hat{\sigma}_z)^T$ is the vector operator whose components are Pauli operators. Then, $\xi_\alpha$ of (\ref{eq:df_xi}) reads
\begin{eqnarray}
\xi_\alpha = \frac{1}{2} \left( 1 + \boldsymbol{\phi}^T\mathbf{R}_\alpha\boldsymbol{\phi} \right),
\label{eq:B_xi}
\end{eqnarray}
where $\mathbf{R}_\alpha$ is a $3 \times 3$ rotation matrix in $\mathbb{R}^3$, whose elements $[\boldsymbol{R}_\alpha]_{jk}$ are given by $[\boldsymbol{R}_\alpha]_{jk} =\frac{1}{2}\Tr{(\hat{X}_\alpha \hat{\sigma}_j \hat{X}_\alpha^\dagger \hat{\sigma}_k)}$ for $j,k = x,y,z$. The rotation angles $\theta_\alpha$ and axes $\mathbf{n}_\alpha$ of $\mathbf{R}_\alpha$ are obtained of a single-qubit unitary operation: 
\begin{eqnarray}
\hat{X}_\alpha = \exp \left[-i\frac{\theta_\alpha}{2}(\mathbf{n}_\alpha^T \boldsymbol{\sigma})\right] = \cos{\frac{\theta_\alpha}{2}}\,\hat{\openone}_2 - i \sin{\frac{\theta_\alpha}{2}} \mathbf{n}_\alpha^T \boldsymbol{\sigma}.
\end{eqnarray}
Then, one has the average fidelity $F$:
\begin{eqnarray}
F = \frac{p}{4}\sum_{\alpha=0}^3 \frac{1}{2}\left[1 + \int d\boldsymbol{\phi} \left( \boldsymbol{\phi}^T\mathbf{R}_\alpha\boldsymbol{\phi} \right) \right]  + \frac{1}{2}\left(1-p\right),
\label{eq:rwF_1q}
\end{eqnarray}
where $d\boldsymbol{\phi}$ is Haar measure over the surface of the Bloch sphere, normalized as $\int d\boldsymbol{\phi}=1$. The integral term is able to be calculated by using the Schur's lemma in $\mathbb{R}^3$ as
\begin{eqnarray}
\int_G dg \, \mathbf{O}_g \mathbf{X} \mathbf{O}_g^T = \frac{1}{r}\Tr{(\mathbf{X})}\,\mathbf{I}_{r},
\label{eq:Bschur}
\end{eqnarray}
where $\mathbf{I}_r$ is identity matrix in $\mathbb{R}^r$, $\mathbf{O}_g$ is an irreducible orthogonal representation of an element $g \in G$, and $dg$ is the Haar measure. This holds for every matrix $\mathbf{X}$ on $\mathbb{R}^r$. By using the Schur's lemma, we have
\begin{eqnarray}
\int d\boldsymbol{\phi} (\boldsymbol{\phi}^T\mathbf{R}_\alpha\boldsymbol{\phi}) = \frac{1}{3}\Tr{(\mathbf{R}_\alpha)},
\label{eq:intB_xi}
\end{eqnarray}
and finally the average fidelity reads
\begin{eqnarray}
F = \frac{1}{2} + \frac{p}{24}\sum_{\alpha=0}^{3} \Tr{(\mathbf{R}_\alpha)}.
\label{eq:fF_d2}
\end{eqnarray}
Noting that $-1 \le \Tr{(\mathbf{R}_\alpha)} = 2\cos\theta_\alpha + 1 \le 3$, one has
\begin{eqnarray}
\left[ F_\text{min}=\frac{1}{2}\left(1 -\frac{p}{3}\right) \right] \le F \le \left[ F_\text{max}=\frac{1}{2} \left( 1+p \right) \right].
\label{eq:Fmin_Fmax}
\end{eqnarray}
Here, the maximum average fidelity $F_\textrm{max}$ is reachable when
\begin{eqnarray}
\hat{X}_0 = \hat{X}_1 = \hat{X}_2 = \hat{X}_3 = \hat{\openone}_2,
\label{eq:optF_condi1}
\end{eqnarray}
or equivalently, 
\begin{eqnarray}
\hat{U}_\alpha = \hat{V}_\alpha,~\textrm{for}~\alpha=0,1,2,3.
\label{eq:optF_condi2}
\end{eqnarray}
This result is consistent with the previous studies \cite{Horodecki99,Albeverio02}.

\subsection{Fidelity deviation $D$}
Using the formulas of (\ref{eq:f_d}), (\ref{eq:B_xi}), and (\ref{eq:rwF_1q}), we can represent the fidelity deviation as
\begin{eqnarray}
D = \sqrt{\int d\phi \, f^2 - F^2} = \frac{p}{4}\sqrt{\sum_{\alpha,\beta=0}^{3} c_{\alpha\beta}},
\label{eq:D_d}
\end{eqnarray}
where $c_{\alpha\beta}$ are elements of $4 \times 4$ covariance matrix $\mathbf{C}$, 
\begin{eqnarray}
c_{\alpha\beta} = \int d\phi \, \xi_\alpha \xi_\beta  - \int d\phi \, \xi_\alpha \int d\phi \, \xi_\beta.
\end{eqnarray}
Note that $\mathbf{C}$ is symmetric, i.e., $c_{\alpha\beta} = c_{\beta\alpha}$, and its diagonal elements $c_{\alpha\alpha}$ are equal to 
\begin{eqnarray}
\delta_\alpha^2 = \int d\phi \, \xi_\alpha^2 - \left(\int d\phi \, \xi_\alpha\right)^2.
\label{eq:delta_xi}
\end{eqnarray} 
Each element $c_{\alpha\beta}$ is bounded as 
\begin{eqnarray}
-\frac{1}{2}\delta_\alpha \delta_\beta \le c_{\alpha\beta}\le \delta_\alpha \delta_\beta ~(\alpha \neq \beta),
\label{eq:ineq_Cd2}
\end{eqnarray}
where the lower bound is saturated when the two rotation axes $\mathbf{n}_\alpha$ and $\mathbf{n}_\beta$ are orthogonal to each other, and the upper bound is when $\mathbf{n}_\alpha$ and $\mathbf{n}_\beta$ are parallel or antiparallel (See Appendix B in \cite{Bang12} for the details). Then, the fidelity deviation $D$ is lower bounded as
\begin{eqnarray}
D \ge \frac{p}{4}\sqrt{\sum_{\alpha=0} ^{3} \delta_\alpha^2 - \frac{1}{2}\sum_{\alpha \neq \beta}^3 \delta_\alpha \delta_\beta}.
\end{eqnarray}
Here, note that the condition (\ref{eq:optF_condi1}) [or (\ref{eq:optF_condi2})] is a sufficient condition to the zero fidelity deviation. The condition (\ref{eq:optF_condi1}) can be rewritten as $\xi_\alpha = 1$ ($\forall \alpha$) [refer to (\ref{eq:df_xi})]. Then, all elements of the covariance matrix $\mathbf{C}$ become zero, and it leads to zero fidelity deviation by (\ref{eq:D_d}). In other words, maximizing average fidelity $F$ automatically guarantees the minimum of fidelity deviation (even when $F < 1$). This holds for arbitrary noise parameter $p$. However, the inverse is not trivial, i.e., the zero fidelity deviation does not guarantee the maximizing average fidelity. We shall see this later in Fig. \ref{fig:telep_spd2}. Then, the upper bound of $D$ is given as
\begin{eqnarray}
D \le \frac{p}{4}\sqrt{\sum_{\alpha=0} ^{3} \delta_\alpha^2 + \sum_{\alpha \neq \beta} \delta_\alpha \delta_\beta} = \frac{p}{4}\sum_{\alpha=0}^{3}\delta_\alpha.
\label{eq:rw2D_1q}
\end{eqnarray}
The maximum is obtained by the following. By using (\ref{eq:B_xi}), (\ref{eq:intB_xi}), and (\ref{eq:delta_xi}), we rewrite $\delta_\alpha$ in the Bloch representation:
\begin{eqnarray}
\delta_\alpha = \frac{1}{2} \sqrt{\int d\boldsymbol{\phi}\, \left(\boldsymbol{\phi}^T\mathbf{R}_\alpha\boldsymbol{\phi}\right)^2  - \frac{1}{9}\Tr{(\mathbf{R}_\alpha)}^2}.
\label{eq:wrB_delta_xi}
\end{eqnarray}
The integration can be calculated by using Schur's lemma to the product of the two real vector spaces $\mathbb{R}^r \otimes \mathbb{R}^r$,
\begin{eqnarray}
\int_G dg \left(\mathbf{O}_g^T\otimes\mathbf{O}_g^T\right)\mathbf{X}\left(\mathbf{O}_g\otimes\mathbf{O}_g\right) = a \mathbf{I}_{d^2} + b \mathbf{D} + c \mathbf{P}, \nonumber
\label{eq:Bschur2}
\end{eqnarray}
where
\begin{eqnarray}
a &=& \frac{(r+1)\Tr{(\mathbf{X})} - \Tr{(\mathbf{X}\mathbf{D})} - \Tr{(\mathbf{X}\mathbf{P})}}{r(r-1)(r+2)}, \nonumber \\
b &=& \frac{-\Tr{(\mathbf{X})} + (r+1)\Tr{(\mathbf{X}\mathbf{D})} - \Tr{(\mathbf{X}\mathbf{P})}}{r(r-1)(r+2)}, \nonumber \\
c &=& \frac{-\Tr{(\mathbf{X})} - \Tr{(\mathbf{X}\mathbf{D})} + (r+1)\Tr{(\mathbf{X}\mathbf{P})}}{r(r-1)(r+2)}. \nonumber
\end{eqnarray}
Here, $\mathbf{P}$ is a swap matrix such that $\mathbf{P}\,(\mathbf{x}_i \otimes \mathbf{x}_j) = \mathbf{x}_j \otimes \mathbf{x}_i$, or equivalently
\begin{eqnarray}
\mathbf{P}=\sum_{i,j=0}^{r-1}\left(\mathbf{x}_j\otimes\mathbf{x}_i \right) \left(\mathbf{x}_i\otimes\mathbf{x}_j \right)^T. \nonumber
\end{eqnarray}
And 
\begin{eqnarray}
\mathbf{D}=\left(\sum_{i=0}^{r-1} \mathbf{x}_i\otimes\mathbf{x}_i \right) \left( \sum_{j=0}^{r-1}\mathbf{x}_j\otimes\mathbf{x}_j \right)^T, \nonumber
\end{eqnarray}
where $\{\mathbf{x}_i\}$ is an orthonormal basis set in $\mathbb{R}^r$. Using the lemma (\ref{eq:Bschur2}), we have
\begin{eqnarray}
\delta_\alpha = \frac{1}{2\sqrt{5}}-\frac{1}{6\sqrt{5}}\Tr{(\mathbf{R}_\alpha)},
\end{eqnarray}
where we used the following properties: 
\begin{eqnarray}
&& \Tr{(\mathbf{R}_\alpha\otimes\mathbf{R}_\alpha)} = \Tr{(\mathbf{R}_\alpha)}^2,  \nonumber \\
&& \Tr{(\mathbf{R}_\alpha\otimes\mathbf{R}_\alpha\,\mathbf{D})} = \Tr{(\mathbf{R}_\alpha\mathbf{R}_\alpha^T)}=\Tr{(\mathbf{I}_3)}=3,  \nonumber \\
&& \Tr{(\mathbf{R}_\alpha\otimes\mathbf{R}_\alpha\,\mathbf{P})} = \Tr{(\mathbf{R}_\alpha^2)}. \nonumber
\end{eqnarray}
Finally we have the upper bound of $D$ as
\begin{eqnarray}
D \le \frac{1}{\sqrt{5}} \left( F_\textrm{max} - F \right),
\label{eq:upb_d2}
\end{eqnarray}
which only depends on the average fidelity $F$. As we pointed out before, one can see that the zero fidelity deviation is derived when $F=F_\text{max}$.

\subsection{Characterization on the space of ($F$, $D$)}

\begin{figure}[t]
\centering
\includegraphics[width=0.7\textwidth]{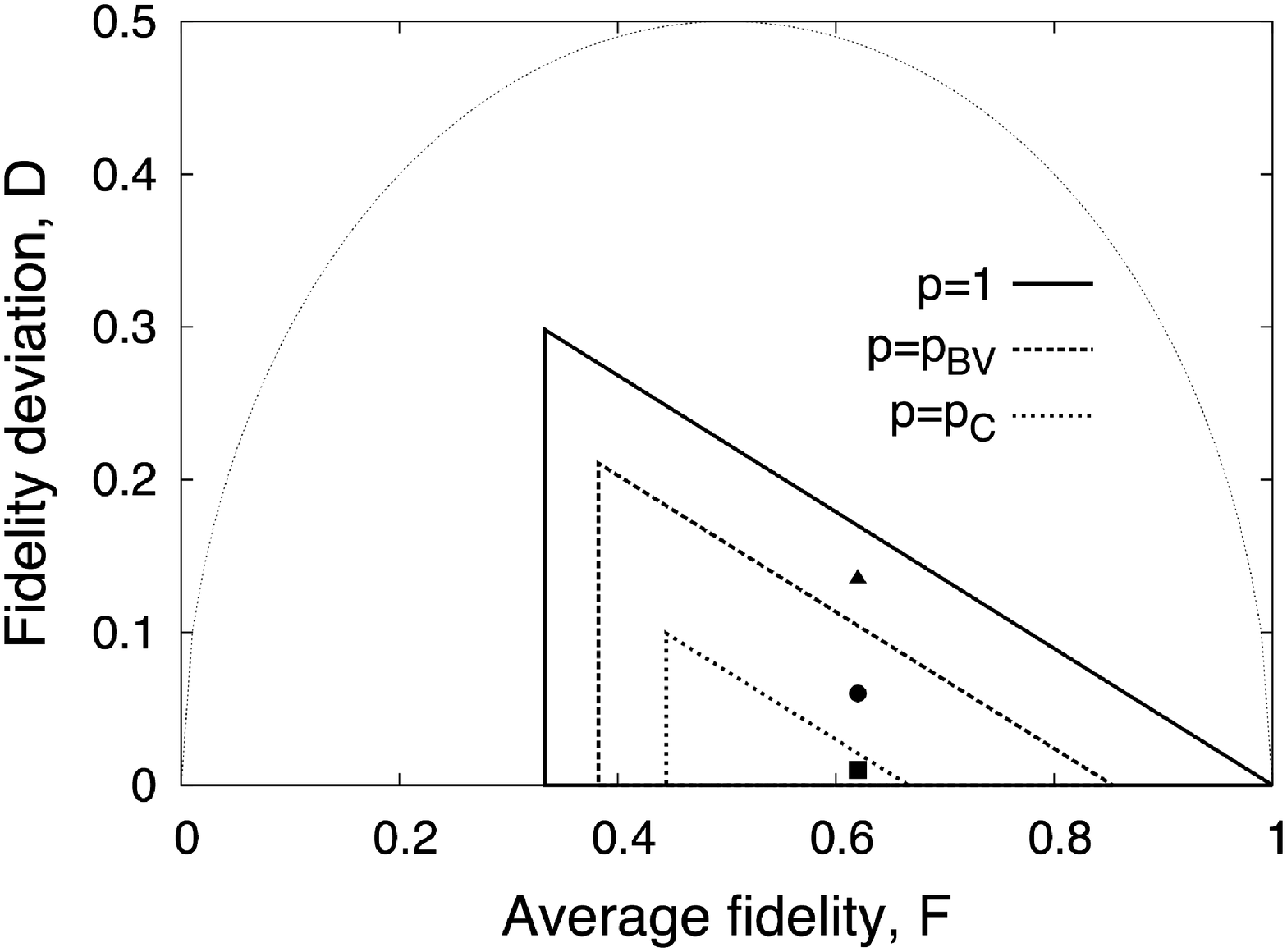}
\caption{Two-dimensional space of the average fidelity and the fidelity deviation. The two performance measures are depicted as a point in this space. The large half circle is the graph of the formula of $D = \sqrt{F-F^2}$, which indicates a mathematical bound in (\ref{eq:limit_ug}). That is, any points (either from classical or quantum) should be placed inside the circle. We present the three different triangular regions characterized by $p=1$ (solid line), $p=p_\textrm{BV}=1/\sqrt{2}$ (dashed line), and $p=p_\textrm{C}=1/3$ (dotted line). Note that $p > p_\text{C}$ is the condition for $F_\text{max} > 2/3$. The three points, denoted by $\blacksquare$, $\CIRCLE$, and $\blacktriangle$, indicate the same average fidelity $F$ but different fidelity deviation $D$ (see the main text for details).}
\label{fig:telep_spd2}
\end{figure}

To characterize our analysis, we here introduce a two-dimensional space of the average fidelity and the fidelity deviation ($F$, $D$), in which the performances of the teleportation are represented as a point. The possible points for the given noise parameter $p$ are localized inside a triangular region. From the conditions of two measures in (\ref{eq:Fmin_Fmax}) and (\ref{eq:upb_d2}), one can construct the triangular region for the following vertices:
\begin{eqnarray}
(F_\text{max}, 0),~(F_\text{min}, 0), ~\text{and}~ (F_\text{min}, D_\text{max}), 
\label{eq:extremal_p}
\end{eqnarray}
where $D_\text{max}=(F_\text{max} - F_\text{min})/\sqrt{5}=2 p/3\sqrt{5}$ [refer to (\ref{eq:upb_d2})]. Each region shows an achievable bound for the teleportation with the given noisy quantum channel, which is determined by the average fidelity $F$ and fidelity deviation $D$. In other words, the size and position of the regions depend on the noise parameter $p$. Here, the extremal points in~(\ref{eq:extremal_p}) are explained in the following situations: For arbitrary $p$, maximizing the average fidelity gives rise to the zero fidelity deviation, i.e., $(F_\text{max}, 0)$ when the condition (\ref{eq:optF_condi1}) [or (\ref{eq:optF_condi2})] holds. However, it is impractical to achieve such an optimal case due to the unexpected noises in the operations in Alice and Bob's sides. Indeed, one can observe the worst case scenario of ($F_\text{min}$, $D_\text{max}$) when the measurement results $m_\alpha$ (a classical signal Alice would send to Bob, see the first paragraph in Sec. \ref{SEC:3}) are flipped, i.e., $m_i \rightarrow m_{j \neq i}$ ($i,j=0,1,2,3$), during the transmission through the classical channel.


For further study, we specify some important regions, each of which is drawn by the following condition (see Fig. \ref{fig:telep_spd2}): (a) $p=1$ (solid line), (b) $p=p_\text{BV}=1/\sqrt{2}$ (dashed line), and (c) $p=p_\text{C}=1/3$ (dotted line). Here, the value $p_\text{BV}$ is the critical value of the Werner state allowing the violation of Clauser-Horne-Shimony-Holt (CHSH) inequality \cite{Clauser69}, and the $p_{\text{C}}$ is of the separability~\cite{Werner89}. Here, $p > p_\textrm{C}$ is known as the condition for $F_\textrm{max}>2/3$, which can never be obtained by any classical strategies. The regions specified by these different conditions are also suggestive of the previously hypothesized result---the dissimilar aspects of the entanglement in the teleportation fidelity and the CHSH inequality violation (together, see Ref.~\cite{Werner89, Popescu94}). However, noting that we can find the characterization points ($F$, $D$) that have same average fidelity but different fidelity deviations, it appears to be more informative to characterize the teleportation performances using both $F$ and $D$. For example, see the three points---square ($\blacksquare$), circle ($\CIRCLE$), and triangle ($\blacktriangle$) in Fig.~\ref{fig:telep_spd2}---each of which is located in the different channel regions.

\section{Summary and closing remarks}

We have analyzed the two different measures, average fidelity $F$ and fidelity deviation $D$, in the teleportation with the noisy quantum channel. In our analysis, we found the conditions to optimize the measures, such that the average fidelity $F$ is maximized and the fidelity deviation $D$ is minimized. We characterized the performances of the teleportation as points in the two-dimensional space of the measures ($F$, $D$). The points could be placed in the triangular regions whose size and position are determined by the noise parameter $p$. Through further analysis, we specified some triangular regions for the different channel conditions, in which we could argue the dissimilar aspects of the entanglement in the teleportation and the violation of CHSH inequality. We hope that our analysis of the fidelity deviation will be applied to other tasks, for example, quantum cloning~\cite{Buzek96}, entangling power~\cite{Zanardi00,Paternostro06}.

\ack
We acknowledge the financial support of the Basic Science Research Program through the National Research Foundation of Korea (NRF) funded by the Ministry of Science, ICT \& Future Planning (No. 2014R1A2A1A10050117) and the National Research Foundation, Prime Minister’s Office, Singapore and the Ministry of Education, Singapore under the Research Centres of Excellence programme. D.K. was supported by the National Research Foundation and Ministry of Education in Singapore.



\end{document}